%% file: siam.tex
\documentclass[final]{siamltex}
\usepackage{geometry}  
\newdimen\tempdimen  
\newbox\tempbox
\newdimen\tabledim

\def\tablefont{\fontsize{8}{9}\selectfont}%
\def\tabnotefont{\fontsize{8}{9}\selectfont}

\long\def\tbl#1#2{%
\setbox\tempbox\hbox{\tablefont #2}%
\tabledim\hsize\advance\tabledim by -\wd\tempbox
\tempdimen\wd\tempbox
\global\divide\tabledim2
\caption{#1\protect\vphantom{yp}}
\centerline{\box\tempbox}}%

\makeatletter
\newenvironment{tabnote}{%
\par\vskip5pt
\tabnotefont
\@ifnextchar[{\@tabnote}{\@tabnote[]}}{%
\par}
\def\@tabnote[#1]{\def\@Tempa{#1}\leftskip\tabledim\rightskip\leftskip\ifx\@Tempa\@empty\else{\it #1:}\ \fi\ignorespaces}
\makeatother

\def\Note#1#2{\parindent0pt\par{\it #1}\ #2}
\input{preamble.tex}

\graphicspath{{figures//}}

\title{Finite Element Integration with Quadrature on the GPU}
\author{Matthew G. Knepley \footnotemark[1]
  \and Karl Rupp \footnotemark[2]
  \and Andy Terrel \footnotemark[3]}

\begin{document}
\maketitle

\pagestyle{myheadings}
\thispagestyle{plain}
\markboth{KNEPELY, RUPP, TERREL}{FEM QUADRATURE ON THE GPU} 

\renewcommand{\thefootnote}{\fnsymbol{footnote}}

\footnotetext[1]{Department of Computational and Applied Mathematics, Rice University, Houston, TX
  (knepley@rice.edu). Supported by the Office of Advanced Scientific Computing Research, Office of Science,
  U.S. Department of Energy, under Contract DE-AC02-06CH11357 and NSF Grant NSF SI2-SSI 1450339.}

\footnotetext[2]{Freelance Scientist, Vienna, AT (me@karlrupp.net).}

\footnotetext[3]{Fashion Metric, Austin, TX (andy.terrel@gmail.com)}

\renewcommand{\thefootnote}{\arabic{footnote}}

\input{abstract.tex}



\input{document.tex}

\bibliographystyle{siam} 
\bibliography{siam}
\end{document}

%% file: preamble.tex
\usepackage[T1]{fontenc}
\usepackage{amsmath,amssymb,amsfonts,graphicx}
\usepackage{epsfig}
\usepackage{epstopdf}
\usepackage{subfigure, float}
\usepackage{algorithmic}
\usepackage{algorithm}
\usepackage[parfill]{parskip}
\usepackage{comment}
\usepackage{cite}
\usepackage{subfloat}
\usepackage{listings}
\usepackage{alltt}
\usepackage[usenames,dvipsnames]{xcolor}
\usepackage{hyperref}

\usepackage{tikz}
\usepackage{pgflibraryshapes}
\usetikzlibrary{decorations.pathreplacing}
\usetikzlibrary{backgrounds}
\usetikzlibrary{arrows}

\def\vf{{\bf f}}

%% file: abstract.tex
\begin{abstract}
  We present a novel, quadrature-based finite element integration method for low-order elements on GPUs, using a pattern
  we call \textit{thread transposition} to avoid reductions while vectorizing aggressively. On the NVIDIA
  GTX580, which has a nominal single precision peak flop rate of 1.5 TF/s and a memory bandwidth of 192 GB/s, we achieve close to 300 GF/s for element integration on first-order discretization of the Laplacian operator with variable coefficients in two dimensions, and over 400 GF/s in three dimensions.
  From our performance model we find that this corresponds to 90\% of our measured achievable bandwidth peak of 310 GF/s.
  Further experimental results also match the predicted performance when used with double precision (120 GF/s in two dimensions, 150 GF/s in three dimensions).
  Results obtained for the linear elasticity equations (220 GF/s and 70 GF/s in two dimensions, 180 GF/s and 60 GF/s in three dimensions) also demonstrate the applicability of our method to vector-valued partial differential equations.
\end{abstract}

%% file: document.tex
\lstset{language=c,
  basicstyle=\fontfamily{cm}\normalfont,
  keywordstyle=\color{ForestGreen}\bfseries,
  %identifierstyle=\color{brown},
  morekeywords={float2,float3,float4},
  emph={__global__,__shared__},
  emphstyle=\color{Mulberry}\bfseries,
  emph={[2]integrateIdentityAction},
  emphstyle={[2]\color{RoyalBlue}\bfseries},
  texcl=true,
  commentstyle=\color{Red},
}

%STREAM from ViennaCL
%NVIDIA GTX 285:    134 GB/s (159)
%NVIDIA GTX 580:    166 GB/s (192)
%AMD HD7970:        199 GB/s (288)
%Dual Intel E5-2670: 80 GB/sec (101)
%Intel Xeon Phi:     95 GB/sec (220/320)

\section{Introduction}
\label{sec:intro}

Despite the large body of research on finite element methods for accelerators, widely-used, freely available general
finite element codes and libraries do not use them, although exceptions exist~\cite{MindenSmithKnepley2013,AbaqusManual}.  This trend is troubling as a growing number of high performance machines rely
on accelerator technologies for the majority of their performance~\cite{Top500}.  Most research is limited to high-order methods~\cite{KlocknerWarburtonBridgeHesthaven2009,KomatitschErlebacherGoddekeMichea2010}
or one-off specialized solvers.  In particular, the first and third author's previous work focused on a specialized
version of the finite element integration routines~\cite{KnepleyTerrel2013}.
Providing a general, low-order integration method for the GPU based on quadrature will be necessary for the wider use of these architectures by finite
element codes.

Our goal is to provide a general interface for efficient evaluation of finite element integrals on the limited memory,
highly concurrent architectures of graphics processing units (GPUs).
To support a wide set of application codes, we are
interested in weak forms incorporating complex coefficients, even those which cannot be accurately represented in the
finite element basis. These forms necessitate the use of quadrature for integration, and are thus more widely applicable
than methods involving exact integration. Moreover, quadrature may even be applicable in cases where coefficients can be
represented in the finite element basis since the size of tensors produced by exact integration grows
quickly~\cite{Ølgaard2010}.

Even though low-order finite element methods are the most used in application codes, no quadrature-based efficient
mapping to accelerator technologies exists. A popular choice for concurrent integration of finite element weak forms is
to assign each cell to a separate thread~\cite{CorriganCamelliLohnerWallin2011,Cecka2011,TaylorChengOurselin2008,Williams12,DziekonskiSypekLameckiMrozowski2012}. This strategy, however,
uses a large amount of local memory per thread. An alternative strategy~\cite{MILAMIN08} vectorizes the computation over
basis functions, taking each quadrature point in turn. This requires very little local memory, but pulls all finite
element coefficients from global memory for each quadrature point, rather than a single time, which is suboptimal for
bandwidth-limited computations.

The method developed in this paper uses a \textit{thread transposition} operation to map between evaluation at
quadrature points for integration and on basis elements. This method increases the size of local memory usage slightly
but avoids synchronization points. With the removal of the synchronization points and increased concurrency, the
implementation achieves higher performance. In addition, it is able to hide much of the latency of moving data to local
memory by keeping many kernels in flight on each processing element. In order to evaluate our method, we developed a
model of both the memory traffic and computation, which allows us to predict performance based upon problem parameters.

In this paper, we formulate a simple interface that uses quadrature, supports low-order elements efficiently, and
achieves 90\% of memory bandwidth peak, up to 300 GFlops, on a variety of accelerators. This verifies our comprehensive
performance model which is able to predict the performance within 10\%. This novel execution strategy
gives the integration method a considerable boost in performance, but more importantly the concision and simplicity of
the implementation greatly reduces the effort of porting to different architectures, as well as the effort for inclusion
in an existing code base.  This contribution provides a road forward for widely used finite element methods to use
accelerators effectively without large refactoring and reformulation efforts.

\section{Formulation}
\label{sec:formulation}

If we restrict our attention to weak forms dependent only on problem fields and their gradients, we can formulate a
generic scalar weak form as~\cite{brown2010ens}
\begin{equation}\label{eq:weakForm}
  \int_\Omega \phi\cdot f_0(u,\nabla u) + \nabla\phi:\vf_1(u,\nabla u) = 0.
\end{equation}
We extend this to vector forms merely by making $f_0$ a vector and $f_1$ a tensor. Breaking the integral into element
integrals, and using quadrature for element integrals, we have
\begin{equation}\label{eq:weakFormDiscrete}
  \sum_e \mathcal{E}^T_e \left[ B^T W f_0(u^q, \nabla u^q) + \sum_k D^T_k W \vf^k_1(u^q, \nabla u^q) \right] = 0
\end{equation}
where $u^q$ is the vector of field evaluations over the set of quadrature points, $W$ a diagonal matrix of quadrature
weights, and the $B$ and $D_k$ matrices implement the reduction over quadrature points to produce basis function
coefficients which are assembled into the global vector by $\mathcal{E}$. The functions $f_0,\vf_1$ encapsulate the
physics of the problem and need only be evaluated pointwise, at quadrature points. Note that these functions will
contain any variable cofficients of your equations, as well as the basic operators. For example, a standard Laplacian
problem uses $\vf_1 = \nabla u$, or $\vf_1 = \kappa(u, x) \nabla u$ for a variable coefficient problem.

% Old Idea
A popular strategy on multi-core processors as well as uniprocessors with vector instructions is to vectorize over the operations used in
integration so that several threads are used for each element~\cite{brown2010ens}. One can vectorize over quadrature points, which is
natural since users typically specify a pointwise ``physics'' function which operates at quadrature points, as shown
above. However, a reduction over quadrature points is necessary in order to calculate the coefficient for each basis
function. This reduction introduces synchronization, which will destroy most of the gains on a GPU coming from
vectorization. It is also possible to vectorize over basis functions, but this strategy introduces both redundant
computation and data movement.

% New Idea
To mitigate the synchronization penalty, we would like the reduction to be calculated by a single thread, instead of
reducing across threads in a block. For this to work, we need enough reductions for all our threads, so we must use
multiple elements per thread block. In order to satisfy Little's Law~\cite{Little61}, this means we also need to use
multiple elements in our quadrature point vectorization.

In fact, the number of elements can be the least common multiple of the number of quadrature points $N_{\mathrm{q}}$ and the number
of (scalar) basis functions per element $N_{\mathrm{b}}$, $\mathrm{LCM}(N_{\mathrm{q}}, N_{\mathrm{b}})$. This is also the number of threads for each thread block
that will be completely occupied. This situation is illustrated in Fig.~\ref{fig:threadTranspose}, which will be
detailed below. In the first phase, we vectorize over quadrature points, and in a second phase, we vectorize over basis
functions. This strategy means that we must keep the intermediate results in shared memory.

\section{Thread Transposition}\label{sec:transpose}

During residual evaluation using quadrature, we would like to spread the evaluation of the residual contribution at each
quadrature point over separate threads, but we also would like a separate thread to evaluate each basis coefficient of
the residual. These two strategies are at odds when we have different numbers of quadrature points, $N_{\mathrm{q}}$, and basis
functions, $N_{\mathrm{b}}$, per element. We will reconcile them using a pattern we call \textit{thread transposition}, in which we
leave data in shared memory and change the target for each thread in different stages of the algorithm.

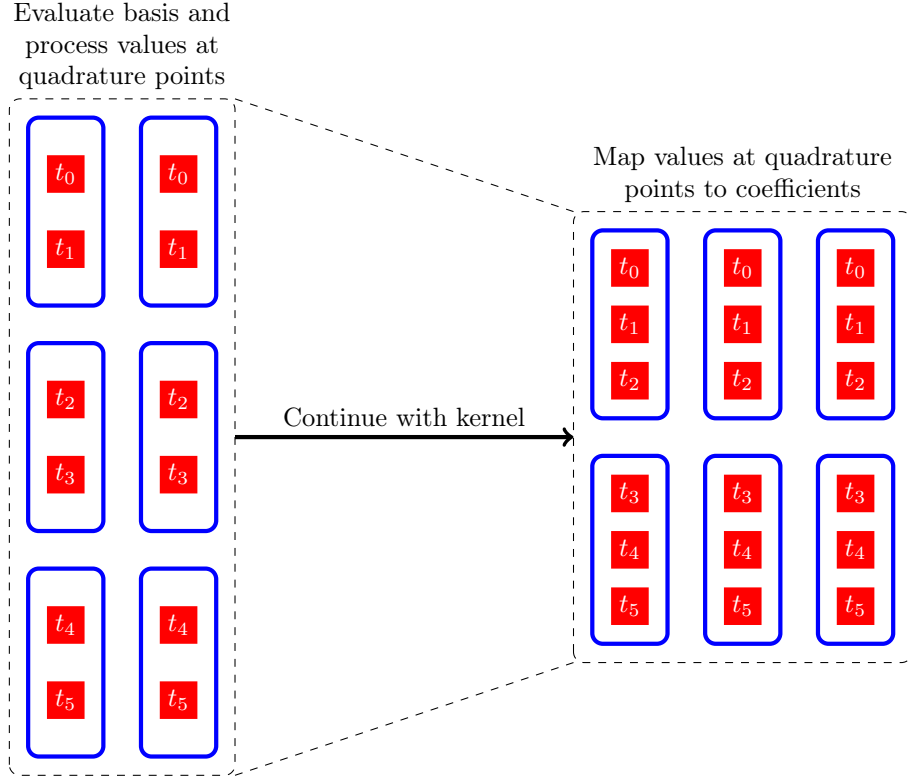
\begin{figure*}[!t]
\centering
\tikzstyle{work group} = [draw,blue,rounded corners,ultra thick]
\tikzstyle{thread}     = [fill,red]
\tikzstyle{operation}  = [dashed,rounded corners]
\tikzstyle{notation}   = [anchor=south,text width=4cm,text centered]
\begin{tikzpicture}[scale=0.5]
% Draw the quadrature point evaluation
\draw[operation] (-0.5,-3.5) rectangle +(6,18)
  ++(3,18) node[notation] {Evaluate basis and process values at quadrature points};
% Draw the quadrature point evaluation breakdown
\foreach \x in {0, 3}
  \foreach \y/\j in {-3/2, 3/1, 9/0}
{
  % draw a cell work unit
  \path[work group] (\x,\y) rectangle +(2,5);
  % draw a 2 thread configuration
 \path[thread] (\x,\y)
             ++(0.5,1.0)
              +(0,0) rectangle +(1,1) +(0.5,0.5) node[anchor=mid,text=white] {\pgfmathtruncatemacro{\t}{\j*2+1}$t_{\t}$}
             ++(0.0,2.0)
              +(0,0) rectangle +(1,1) +(0.5,0.5) node[anchor=mid,text=white] {\pgfmathtruncatemacro{\t}{\j*2+0}$t_{\t}$};
}
% Draw arrow for kernel continuation
\draw[->,ultra thick] (5.5,5.5) -- node[anchor=south] {Continue with kernel} (14.5,5.5);
\draw[dashed] (5.5,14.5) -- (14.5,11.5);
\draw[dashed] (5.5,-3.5) -- (14.5,-0.5);
% Draw the basis function evaluation
\draw[operation] (14.5,-0.5) rectangle +(9,12)
  ++(4.5,12) node[notation] {Map values at quadrature points to coefficients};
% Draw the basis function evaluation breakdown
\foreach \x in {15, 18, 21}
  \foreach \y/\j in {0/1, 6/0}
{
  % draw a cell work unit
  \path[work group] (\x,\y) rectangle +(2,5);
  % draw a 3 thread configuration
 \path[thread] (\x,\y)
             ++(0.5,0.5)
              +(0,0) rectangle +(1,1) +(0.5,0.5) node[anchor=mid,text=white] {\pgfmathtruncatemacro{\t}{\j*3+2}$t_{\t}$}
             ++(0.0,1.5)
              +(0,0) rectangle +(1,1) +(0.5,0.5) node[anchor=mid,text=white] {\pgfmathtruncatemacro{\t}{\j*3+1}$t_{\t}$}
             ++(0.0,1.5)
              +(0,0) rectangle +(1,1) +(0.5,0.5) node[anchor=mid,text=white] {\pgfmathtruncatemacro{\t}{\j*3+0}$t_{\t}$};
}
\end{tikzpicture}
\caption{Action of the residual evaluation kernel on a group of six incoming cells. Each cell is displayed as a blue,
  rounded rectangle occupied by the threads which compute the cell information. Each thread computes its values in
  series, so that thread $t_0$ first computes values at quadrature points for two cells, and then computes basis
  coefficients for three cells.}
\label{fig:threadTranspose}
\end{figure*}

If we take the size of the thread block as the least common multiple of the quadrature and basis sizes, $N_{\mathrm{t}} = \mathrm{LCM}(N_{\mathrm{q}},
N_{\mathrm{b}})N_{\mathrm{comp}}$, where $N_{\mathrm{comp}}$ denotes the number of components for a possibly vector-valued basis, we can use different size cell blocks to make the sizes commensurate. The quadrature computation will use
$N_{\mathrm{t}}/N_{\mathrm{q}}$ groups of $N_{\mathrm{q}}$ threads, each group operating on a series of $N_{\mathrm{t}}/N_{\mathrm{b}}$ cells. Likewise, the basis coefficient
computation will use $N_t/N_b$ groups of $N_{\mathrm{b}}$ threads, each group operating on a series of $N_{\mathrm{t}}/N_{\mathrm{q}}$ cells. This
arrangement is shown in Fig.~\ref{fig:threadTranspose} for a 2D $P_1$ Lagrange element, where we choose the number of basis functions $N_{\mathrm{b}} = 3$, 
the number of quadrature points $N_{\mathrm{q}} = 2$, and consequently $N_{\mathrm{t}} = 6$ as the least common multiple for the thread block size.
Each thread, shown as a red rectangle, computes values for a group of cells, shown as blue rounded
rectangles, in series. So the thread $t_0$ computes basis function evaluations and the weak form function at quadrature
points for two cells, and then computes a basis coefficient for three cells. This organization necessitates that the
computed products be placed in shared memory so that after transposition different threads have access to the
information. The increased local memory usage from transposition is balanced by the increased concurrency, and thus
overall lower per-thread memory requirements. Note also that for multi-component fields, such as vectors, we instead use
merely the total number of basis functions over all components as $N_{\mathrm{bt}} = N_{\mathrm{b}} N_{\mathrm{comp}}$.

At the top level, we divide the mesh cells into \textit{chunks} which are processed serially by the CPU, or in parallel
on the GPU. Each chunk, of size $N_{\mathrm{chunk}}$, is assigned to a thread block, and thus the number of OpenCL workgroups (or the size of the CUDA grid) is equal to
the number of chunks. The total number of cells is $N_{\mathrm{cells}} = (\mathrm{\#\ of\ chunks}) N_{\mathrm{chunk}} + N_{\mathrm{R}}$, where $N_{\mathrm{R}}$ is the number of remainder
cells which are always processed on the CPU. Since $N_{\mathrm{R}}$ is less than the chunk size $N_{\mathrm{chunk}}$, it is vanishingly small compared
to the total number of elements. Each chunk is further divided into \textit{batches} such that there are
$N_{\mathrm{cb}}$ batches, each consisting of $N_{\mathrm{bc}}$ cells, and $N_{\mathrm{chunk}} = N_{\mathrm{cb}} N_{\mathrm{bc}}$. These batches are executed in sequence by a
thread block. To summarize, chunks are assigned to thread blocks, and divided into batches. Each batch is executed
simultaneously by the thread block.

\begin{figure*}[tbp]
\centering
\includegraphics[width=0.8\textwidth]{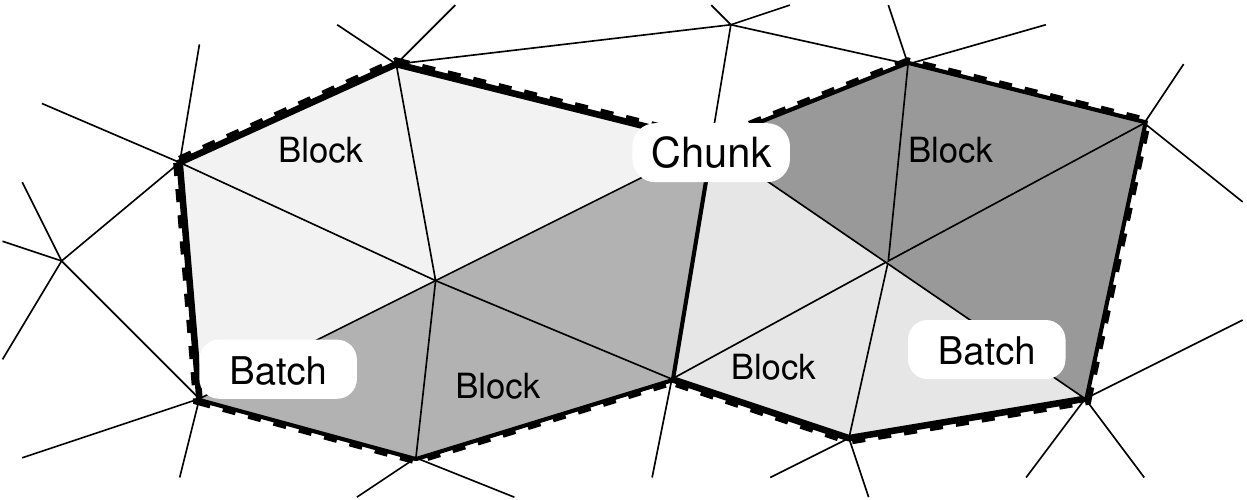}
\caption{Illustration of the top level decomposition:
         The mesh is decomposed into chunks (indicated by bold, dotted lines) assigned to individual thread blocks.
         Each chunk consists of $N_{\mathrm{cb}}$ batches (bold lines), which are processed sequentially.
         Each batch is decomposed into $N_{\mathrm{bl}}$ blocks (shaded regions), which are processed in parallel by the respective thread block.}
\label{fig:chunk-batch-block}
\end{figure*}

Our basic unit of execution, the cell \textit{block}, consists of $N_{\mathrm{bs}} = \mathrm{LCM}(N_{\mathrm{b}}, N_{\mathrm{q}})$ cells, which must divide the batch size
$N_{\mathrm{bc}}$. We execute all blocks in the batch, $N_{\mathrm{bl}} = N_{\mathrm{bc}} /N_{\mathrm{bs}}$, concurrently, but note that all cells in a
block are not processed concurrently. Referring to Fig.~\ref{fig:threadTranspose}, the number of threads in a thread
block will be equal to the block size multiplied by the number of field components and concurrent blocks, $N_{\mathrm{t}} = N_{\mathrm{bs}}
N_{\mathrm{comp}} N_{\mathrm{bl}} = N_{\mathrm{bc}} N_{\mathrm{comp}}$. Each thread first processes $N_{\mathrm{sqc}} = N_{\mathrm{bs}} / N_{\mathrm{b}}$ cells sequentially in the
quadrature phase described in Section~\ref{sec:quadPhase}, which employs $N_{\mathrm{tq}} = N_{\mathrm{q}} N_{\mathrm{comp}}$ threads per
cell. After transposition, each thread processes $N_{\mathrm{sbc}} = N_{\mathrm{bs}} / N_{\mathrm{q}}$ cells sequentially in the basis phase
described in Section~\ref{sec:basisPhase}, which uses $N_{\mathrm{bt}} = N_{\mathrm{b}} N_{\mathrm{comp}}$ threads per cell.

\begin{figure*}[!t]
\centering
\renewcommand{\arraystretch}{1.5}
\tikzstyle{work group} = [draw,blue,rounded corners,ultra thick]
\tikzstyle{thread}     = [red,dashed,rounded corners,ultra thick]
\tikzstyle{batch}      = [olive,rounded corners,ultra thick]
\tikzstyle{block}      = [purple,rounded corners,ultra thick]
\tikzstyle{seq}        = [brown,thick,decorate,decoration={brace,mirror}]
\tikzstyle{seq2}       = [brown,thick,decorate,decoration={brace}]
\tikzstyle{notation}   = [anchor=south,text width=4cm,text centered]
\begin{tikzpicture}[scale=0.333]
% Draw the basis function evaluation breakdown
\path (19, 22.2) node[anchor=south,text centered] {\bf Basis Phase};
\foreach \x in {15, 18, 21}
  \foreach \y in {-12, -6, 1, 7}
{
  % draw a cell work unit
  \path[work group] (\x,\y) rectangle +(2,5) +(1, 2.5) node[anchor=mid,text=black] {$\begin{tabular}{@{}c@{}c@{}} T & T\\T & T\\T & T\end{tabular}$};
}
% Draw the quadrature point evaluation breakdown
\path (2.5, 22) node[anchor=south,text centered] {\bf Quadrature Phase};
\foreach \x in {0, 3}
  \foreach \y in {-18, -12, -6, 1, 7, 13}
{
  % draw a cell work unit
  \path[work group] (\x,\y) rectangle +(2,5) +(1, 2.5) node[anchor=mid,text=black] {$\begin{tabular}{@{}c@{}c@{}} T & T\\T & T\end{tabular}$};
}
% Show number of threads
\draw[thread] (17.5,-12.5) rectangle +(3,25)
  ++(1.5,25) node[notation] {$N_{\mathrm{t}} = 24$};
\draw[thread] (-0.5,-18.5) rectangle +(3,37)
  ++(1.5,37) node[notation] {$N_{\mathrm{t}} = 24$};
% Show number of cells in a batch
\draw[batch] (14,-13) rectangle +(10,27)
  ++(5,27) node[notation] {$N_{\mathrm{bc}} = 12$};
% Show number of cells in a block
\draw[block] (-1,-19) rectangle +(7,19)
  ++(3.5,0) node[notation,anchor=north] {$N_{\mathrm{bs}} = 6$};
% Show number of sequential basis cells
\draw[seq] (15,-14) -- +(8,0) node[notation,pos=0.5,anchor=north] {$N_{\mathrm{sbc}} = 3$};
% Show number of sequential quadrature cells
\draw[seq2] (0,20) -- +(5,0) node[notation,pos=0.5,anchor=south] {$N_{\mathrm{sqc}} = 2$};
% Show the concurrent blocks
\draw (25,6.5) -- ++(3, 0) -- ++(0,-13) node[pos=0.5,anchor=west] {$N_{\mathrm{bl}} = 2$} -- ++(-3,0);
\draw (-2,9.5) -- ++(-3, 0) -- ++(0,-19) node[pos=0.5,anchor=east] {$N_{\mathrm{bl}} = 2$} -- ++(3,0);
\end{tikzpicture}
\caption{Diagram of a single cell batch for a computation similar to Fig.~\ref{fig:threadTranspose} so that $N_{\mathrm{b}} = 3$ and
$N_{\mathrm{q}} = 2$, but now with a vector element with two components ($N_{\mathrm{comp}} = 2$). We choose $N_{\mathrm{bl}} = 2$ concurrent blocks, so that the batch size
is $N_{\mathrm{bc}} = 12$. Each thread is represented by a $T$, and since we have a thread for each component, $N_{\mathrm{t}} = 24$ threads are operating concurrently, with $4$
threads per cell in the quadrature phase, and $6$ threads per cell in the basis phase.}
\label{fig:cellBatch}
\end{figure*}
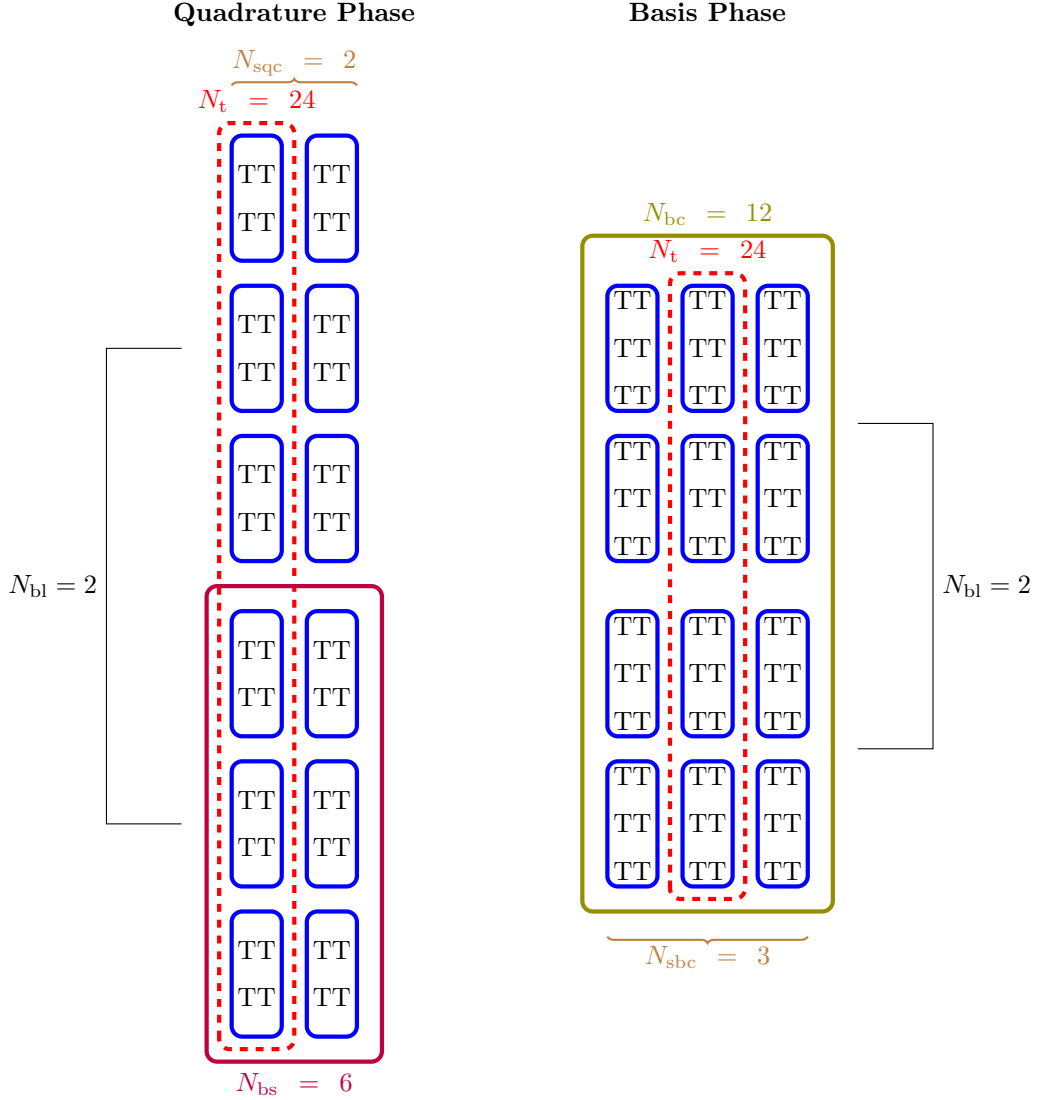

Before loading cell information, we store the quadrature point and weight associated with each thread into thread-private memory,
and load the tabulation of basis functions and derivatives at quadrature points into shared memory. Note that some forms
do not require quadrature points or part of the basis information. We then loop over the batches of cells sent to the
given thread block, first loading geometric information (Jacobian inverse and determinant) and input basis coefficients,
and then executing the two stages discussed below. Note that executing multiple batches in sequence allows computation
in a thread block to overlap the latency for coefficient access in another. The full pseudo-code for the integration
routine is given in the Appendix.

\subsection{Map coefficients to values at quadrature points}\label{sec:quadPhase}
Each thread maps the basis coefficients of $N_{\mathrm{sqc}} =
N_{\mathrm{bs}}/N_{\mathrm{bt}}$ cells to evaluations of the input field and its gradient at the quadrature points. Each thread sums the
contribution from all basis functions at its assigned quadrature point, using the shared Jacobian inverse to transform
the field derivatives. These values are then passed to the $f_0$ and $\vf_1$ functions from Eq.~\ref{eq:weakForm} and
the results, multiplied by the quadrature weight and determinant, are stored in shared memory. After this write, we must
synchronize the thread block in order to make these values available to the other threads, which should take about 20
cycles~\cite{Wong2010}.

\subsection{Map values at quadrature points to coefficients}\label{sec:basisPhase}
After the evaluations at quadrature points have been placed in shared memory and threads have been synchronized, each thread maps the values at quadrature points of
$N_{\mathrm{sbc}} = N_{\mathrm{bst}}/(N_{\mathrm{q}} N_{\mathrm{comp}})$ cells to basis coefficients. A thread forms the product of the $f^k$ values with the
test function and gradient at each quadrature point and accumulates the result in a local variable. Note that this is
the action of the $B^T$ and $D^T_k$ matrices from Eq.~\ref{eq:weakFormDiscrete}. These results are written to global
memory by the thread block for each of the $N_{\mathrm{sbc}}$ cells, which means that $N_{\mathrm{cbc}} = N_{\mathrm{bl}} N_{\mathrm{q}}$ cells are written
concurrently.

\subsection{Memory Traffic and Computation}

The concurrency for our algorithm is at worst $\mathrm{min}(N_{\mathrm{q}}, N_{\mathrm{b}})$. In shared memory, we must store geometric
information (Jacobian inverses and determinants), tabulated basis functions and derivatives, basis function
coefficients, and the $f$ values at quadrature points, so the maximum shared memory used in bytes is given by
\begin{align}
  M = 4 (&(d^2 + 1) N_{\mathrm{t}} + 
               (d + 1) N_{\mathrm{bt}} N_{\mathrm{q}} + N_{\mathrm{t}} N_{\mathrm{bt}} +
               (d + 1) N_{\mathrm{t}} N_{\mathrm{sqc}}),
\end{align}
where $d$ is the spatial dimension, so that per cell we have
\begin{align}
  M_{\mathrm{c}} &= \frac{M}{N_{\mathrm{bc}}} \nonumber\\
      &= 4 N_{\mathrm{comp}} \left((d^2 + 1) + \frac{d + 1}{N_{\mathrm{bl}}} + N_{\mathrm{bt}} + (d + 1) N_{\mathrm{sqc}} \right) \nonumber\\
      &\le 4 N_{\mathrm{comp}} \left((d^2 + 1) + \frac{d + 1}{N_{\mathrm{bl}}} + N_{\mathrm{bt}} + (d + 1) N_{\mathrm{q}} \right)
\end{align}
where we get equality if $\mathrm{LCM}(N_{\mathrm{b}}, N_{\mathrm{q}}) = N_{\mathrm{b}} N_{\mathrm{q}}$. Note that some
of these values can be located in thread-local memory, but this is an upper bound for shared memory. In our examples, we
need only
\begin{equation}
  M_{\mathrm{c}} \le 4 N_{\mathrm{comp}} \left((d^2 + 1) + \frac{d}{N_{\mathrm{bl}}} + N_{\mathrm{bt}} + d N_{\mathrm{q}} \right)
\end{equation}
% 2D P1 Poisson: M_c \le 4*4*1 (5 + 2/N_{bl} + 3 + 2) = 16 (10 + 2/N_{bl}) = 162, M = 162*32 = 5184
since we do not need function tabulation of $f_0$ values. For our Poisson example in Section~\ref{sec:benchmark}, with
blocks of 32 cells, the memory per thread block is $M = 5$ KB. This allows 9 concurrent workgroups to be resident on a
single multiprocessor, allowing overlap of memory access latency by computation from another workgroup. This has been shown to be quite
important, along with the requirement to keep a large number of threads in flight, for optimal performance on the
GPU~\cite{Cruz20112084}.

The memory loaded per batch is
\begin{equation}
  4 N_{\mathrm{t}} \left((d^2 + 1) + N_{\mathrm{bt}} + (d + 1) N_{\mathrm{q}} \right)
\end{equation}
since the tabulation is only loaded once. The computation per cell batch is
\begin{align}
  [((2 + (2 + 2d) d) N_{\mathrm{bt}} + 2 d N_{\mathrm{comp}}) N_{\mathrm{q}} +
                (2 + 2d) d N_{\mathrm{q}} N_{\mathrm{bt}}] N_{\mathrm{bs}} N_{\mathrm{bl}} .
\end{align}
Thus we have for the algorithmic balance expressed in floating point operations per byte
\begin{align}
  \beta &= \frac{\bigl[(2 + (2 + 2d) d) N_{\mathrm{bt}}N_{\mathrm{q}} + 2 d N_{\mathrm{comp}} N_{\mathrm{q}} + (2 + 2d) d N_{\mathrm{q}} N_{\mathrm{bt}} \bigr] N_{\mathrm{bs}} N_{\mathrm{bl}}
                }{4 N_{\mathrm{t}} \left((d^2 + 1) + N_{\mathrm{bt}} + (d + 1) N_{\mathrm{q}} \right)}
        %&= \frac{(1 + 2d + 2d^2) N_{\mathrm{b}} N_{\mathrm{q}} + d N_{\mathrm{q}}}{2 \left((d^2 + 1) + N_{\mathrm{b}} + (d + 1) N_{\mathrm{q}} \right)}
\end{align}
For our example Poisson problem, $d = 2$, $N_{\mathrm{b}} = 3$, $N_{\mathrm{q}} = 1$, and $N_{\mathrm{comp}} = 1$, 
so that
\begin{equation} \label{eq:beta-poisson}
  \beta = \frac{41}{22} \approx 2 \frac{\mathrm{flop}}{\mathrm{byte}},
\end{equation}
hence our computations are limited by memory bandwidth on typical GPUs.
%The bandwidth peak on the Nvidia GTX580 is 192 GB/s, but our STREAMS benchmark
%performance is about 166 GB/s. Thus, we might achieve 310 GF/s, about 20\% of peak at best.

\section{Variable Coefficients}\label{sec:variable-coefficients}

In order to make our proposed algorithm applicable to many applications in computational science and engineering, it must support variable
auxiliary coefficients. These coefficients often come from input data, but can also arise from multiphysics coupling. We
express these coefficients as vectors in a linear space, not necessarily the same finite element space as the solution.

When considering variable coefficients, additional data must be retrieved from global memory and stored in shared memory, which reduces the number of concurrent batches possible on a
single processor. Moreover, if too little flops are executed with the auxiliary field, it can lower $\beta$, our
algorithmic balance.

%GPU memory types\\
%- Can't use textures effectively because no cache reuse\\
%- bandwidth to global memory on write exceeds speedup from texture to sm\\

The auxiliary fields can be accomodated with a small extension to our interface. Our basic equation now becomes
\begin{equation}\label{eq:weakFormVariable}
  \int_\Omega \phi\cdot f_0(u,\nabla u,a,\nabla a) + \nabla\phi:\vf_1(u,\nabla u,a,\nabla a) = 0.
\end{equation}
where $a$ represents the auxiliary fields. On the CPU, our code allows an arbitrary representation for $a$, including
fields supported directly on quadrature points. On the GPU, however, we must match the blocking of $a$ with that of the
solution fields $u$. Our current implementation allows $a$ to come from the same space as $u$, or from the $P_0$ space
of constants over the cell.

\section{GPU Implementation}\label{sec:gpu-implementation}

We derived a first implementation of our algorithm using CUDA.
However, since no CUDA-specific routines such as warp shuffle routines are used, we reimplemented the algorithm using the free, open standard OpenCL and observed the same performance.
Even though OpenCL requires some additional boilerplate code, it provides several advantages when integrated into a library such as PETSc:
First, a broader range of hardware from different vendors can be targeted.
Second, the OpenCL implementation only requires to link with the shared library distributed with the respective OpenCL SDK, hence it is far less intrusive for the build process than CUDA.
Third, the kernels are compiled at runtime by passing the respective kernel source string to the OpenCL just-in-time compiler, which allows for target-specific optimizations at runtime.
Thus, any file-based code generation during the build stage can be avoided and instead the source string can be built entirely from runtime parameters during execution.

We provide default implementations for the Poisson equation and the Lame equation for linear elasticity.
Library users can also provide their own functions for $\vf_0$ and $\vf_1$ defined in \eqref{eq:weakForm}, in which case the source code representing $\vf_0$ and $\vf_1$ defined in \eqref{eq:weakForm} need to be provided as a string.
Because it is more convenient to provide function pointers rather than string representations of functions, this can be considered a disadvantage of using OpenCL.

\section{Benchmarks}\label{sec:benchmark}

\begin{table}
\tbl{Overview of accelerator hardware\label{tab:hardware}}{
 \begin{tabular}{l|ccc|cc}
                                    &  \multicolumn{3}{c|}{NVIDIA}                           & \multicolumn{2}{c}{AMD} \\
                                    &  GTX 580         & Tesla K20m       & GTX 750 Ti       & A10-5800K      & FirePro W9100\\
  \hline
  Peak Mem. BW (GB/sec)             &  \hphantom{1}192 & \hphantom{2}208  & \hphantom{13}88  & \hphantom{1}26 & \hphantom{2}320 \\
  Peak GFLOP/sec (float)            &             1581 &            4106  &            1306  &            614 &            5238 \\
  Peak GFLOP/sec (double)           &  \hphantom{1}198 &            1173  & \hphantom{13}41  &            154 &            2619 \\
  TDP (Watt)                        &  \hphantom{1}244 & \hphantom{1}244  & \hphantom{13}60  &            100 & \hphantom{2}275 \\
 \end{tabular}}
\begin{tabnote}
\Note{Note:}{Values for memory bandwidth and FLOPS are theoretical peaks. (TDP: Thermal Design Power)}
\end{tabnote}
\end{table}

We calculate the residual for the Poisson equation and of a linear elastic problem on unstructured simplex meshes of the unit square in two dimensions and of the unit cube in three dimensions.
The specific benchmark code is distributed with PETSc as SNES example 12, available
at \href{http://www.mcs.anl.gov/petsc/petsc-dev/src/snes/examples/tutorials/ex12.c.html}{http://www.mcs.anl.gov/petsc/petsc-dev/src/snes/examples/tutorials/ex12.c.html}. All
runs can be reproduced using the latest PETSc release and the instructions in Section~\ref{sec:reproduce}. We note that
the OpenCL backend can now be used for all PETSc FEM examples which provide the $f_0$ and $\vf_1$ functions as text. For
the Poisson problem, $f_0$ in \eqref{eq:weakForm} is trivially zero, while for $\vf_1$ we provide the following
function (passed to PETSc as a string) acting pointwise on quadrature points:
\lstset{language=C,basicstyle=\footnotesize,morekeywords={float2}}
\begin{lstlisting}
float2 f1_laplacian(float u[], float2 gradU[], float a[], float2 gradA[], int comp)
{
  return gradU[comp];
}
\end{lstlisting}
For 2D linear elasticity, the function is only slightly more complicated,
\begin{lstlisting}
float2 f1_elasticity(float u[], float2 gradU[], float a[], float2 gradA[], int comp)
{
  float2 f1;

  switch(comp) {
  case 0:
    f1.x = 0.5*(gradU[0].x + gradU[0].x);
    f1.y = 0.5*(gradU[0].y + gradU[1].x);
    break;
  case 1:
    f1.x = 0.5*(gradU[1].x + gradU[0].y);
    f1.y = 0.5*(gradU[1].y + gradU[1].y);
  }
  return f1;
}
\end{lstlisting}
and note that in both cases the function $f_0$ is null since only field gradients are involved in the residual.
These functions are directly inlined into the OpenCL program source code during the source generation step for the OpenCL just-in-time compiler.

The hardware used for the benchmarks is listed in Table~\ref{tab:hardware}.
NVIDIA GPUs have been selected to represent the Fermi (GTX 580), Kepler (K20m) and Maxwell (GTX 750 Ti) architecture generations.
We note that the GTX 750 Ti is limited by firmware to a lower double precision performance of merely 41 GFLOP/sec.
Also, memory error correction is enabled for the Tesla K20m GPU in order to reflect the typical setup in computing centers, while all other GPUs are used without memory error correction.
The AMD GPUs are an integrated GPU (A10-5800K) and a discrete high-end workstation GPU (FirePro W9100).
All benchmarks were run on Linux-based systems, using OpenCL implementations included in the CUDA 6.5 and CUDA 7.0 releases, respectively.

%% Poisson

\begin{table}
\tbl{Performance (GFLOP/sec) obtained for the Poisson equation in two dimensions using single precision (left) and double precision (right) for $66\,049$ unknowns for different numbers of blocks per batch and batches per chunk.
          Overall, only a mild dependence of the performance on the blocks per batch and the number of batches is observed, since the best and worst performance for the parameters considered differs by only $25$ percent.
          \label{tab:poisson-block-batch}}{
\begin{tabular}{|r||r|r|r|r||r|r|r|r|}
  \hline
               & \multicolumn{4}{c||}{Number of Batches} & \multicolumn{4}{c|}{Number of Batches} \\
  %\hline
  Blocks/Batch &     4 &    8 &    12 & 16       & 4 & 8 & 12 & 16 \\
  \hline
  16           &  180 & 181 & 176 & 162          & 85 & 83 & 80 & 81 \\
  20           &  189 & 179 & 177 & 154          & 91 & 88 & 86 & 78 \\
  24           &  183 & 174 & 164 & 164          & 80 & 78 & 76 & 73 \\
  28           &  190 & 193 & 173 & 173          & 81 & 78 & 75 & 70 \\ 
  32           &  209 & 196 & \textbf{209} & 200 & \textbf{102} & 98 & 94 & 88 \\
  36           &  189 & 183 & 173 & 178          & 81 & 78 & 75 & 75 \\
  \hline
 \end{tabular}
}
\end{table}

In Table~\ref{tab:poisson-block-batch} we consider the performance obtained for different numbers of blocks per batch and different numbers of batches per chunk.
Generally, the number of blocks per batch needs to be balanced such that the number of threads per workgroup is large enough, but at the same time shared memory consumption is kept low enough to obtain higher hardware occupancy.
NVIDIA GPUs provide an upper bound of $1024$ threads per workgroup on recent models, while AMD GPUs require workgroup sizes smaller $256$, which can be cast into upper bounds for the number of blocks per batch for a given basis and quadrature.
As the results show, the correct choice of cell block and batch sizes only has a mild influence:
The difference between the best and the worst performance is only $30$ percent, hence good out-of-the-box performance can be provided with suitable default settings.
Also, the performance remains mostly unchanged as the number of batches is changed, which is expected because batches are processed one after another and their number does not influence the amount of shared memory required.

\begin{figure}[tbp]
\centering
  \subfigure[2D, single precision]{\includegraphics[width=0.48\textwidth]{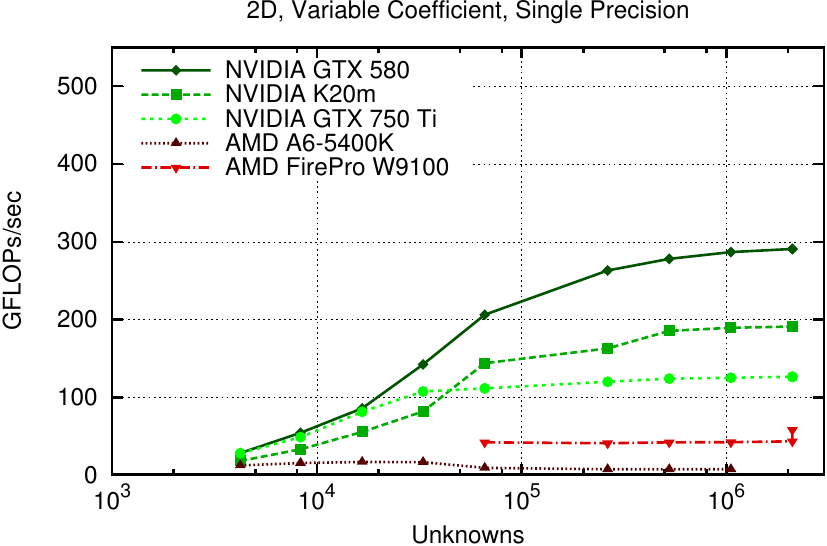}} \hfill
  \subfigure[2D, double precision]{\includegraphics[width=0.48\textwidth]{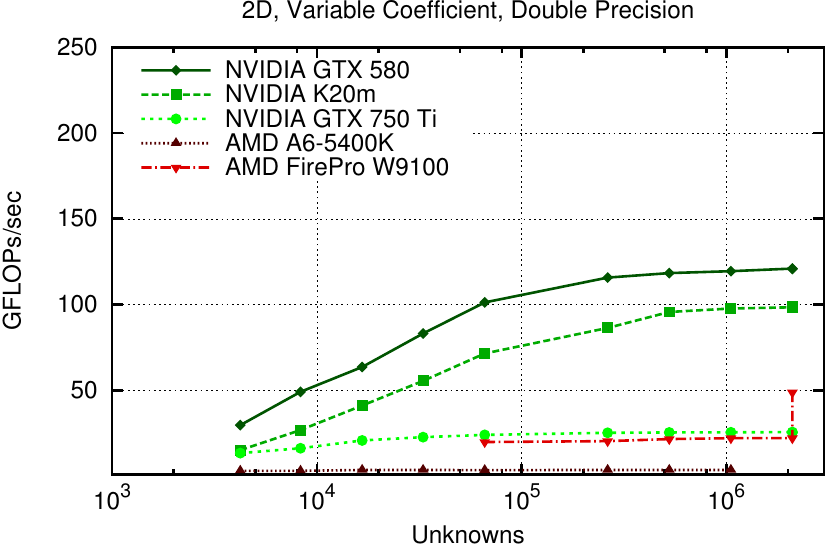}} \\
  
  \subfigure[3D, single precision]{\includegraphics[width=0.48\textwidth]{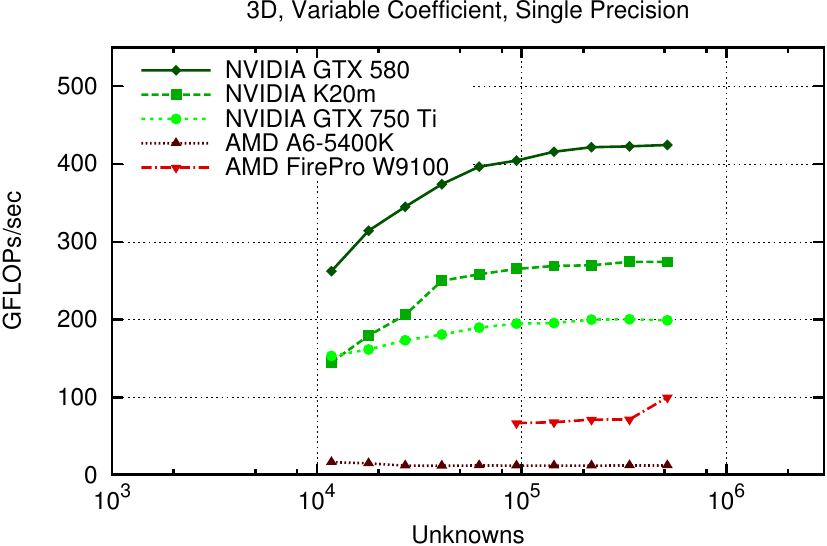}} \hfill
  \subfigure[3D, double precision]{\includegraphics[width=0.48\textwidth]{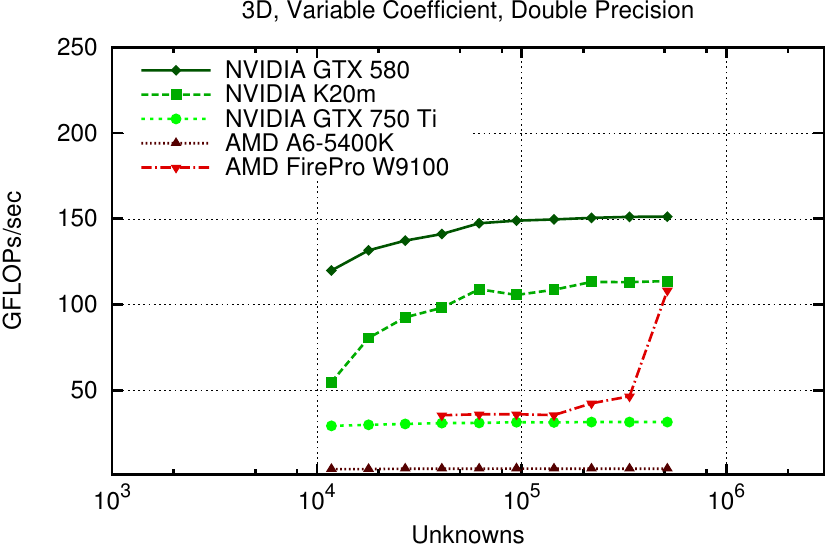}}
\caption{FLOP rates for the different GPUs considered in this benchmark for PETSc SNES example 12, evaluating the residual for the $P_1$ Poisson equation on unstructured simplex meshes of the unit square in two dimensions and on the unit cube in three dimensions.}
\label{fig:flopPoisson}
\end{figure}

Benchmark results obtained for the Poisson problem in Fig.~\ref{fig:flopPoisson} show that on the GTX 580 we obtained up to $420$ GFLOP/sec in single precision for the three-dimensional case.
The $300$ GFLOP/sec in the two-dimensional case amount to an effective memory bandwidth of $150$ GB/sec based on the algorithmic balance $\beta \approx 2 \frac{\mathrm{flop}}{\mathrm{byte}}$ from \eqref{eq:beta-poisson}.
This amounts to $75$ percent of peak memory bandwidth, which is the same as the practical maximum obtained with the STREAM benchmark.
Similarly, the practical peak memory bandwidth is reached for the GTX 750 Ti in single precision, and the firmware limit of double precision performance is reached.
A-priori one expects the Tesla K20m to reach similar performance as the GTX 580, but this is not reflected by the benchmark results.
After closer inspection, however, we could explain the performance difference by the smaller memory bandwidth obtained with the STREAM benchmark, part of which is caused by the memory error correction.

The performance obtained on AMD devices in absolute terms is substantially slower than those we obtained on the devices from NVIDIA.
On closer inspection, however, the integrated A10-5800K GPU achieves a memory bandwidth of $20$ GB/sec, which is the practical maximum in a dual-channel DDR3 configuration.
Only the discrete FirePro W9100 could not reach high bandwidth.
Our investigations showed that this was caused by low effective memory bandwidth achieved when loading the cell coefficients, which are stored in an unstructured way in global memory.

%% Elasticity

\begin{figure}[tbp]
  \centering
  \subfigure[2D, single precision]{\includegraphics[width=0.48\textwidth]{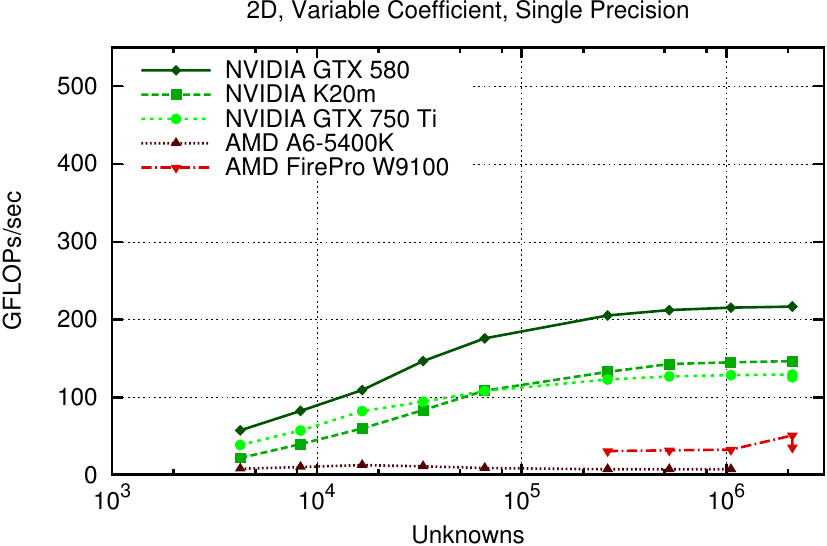}} \hfill
  \subfigure[2D, double precision]{\includegraphics[width=0.48\textwidth]{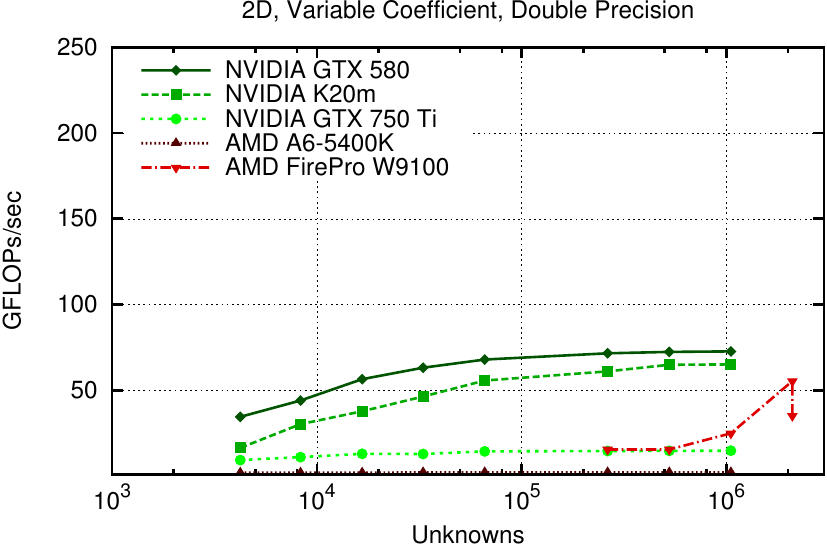}} \\
  
  \subfigure[3D, single precision]{\includegraphics[width=0.48\textwidth]{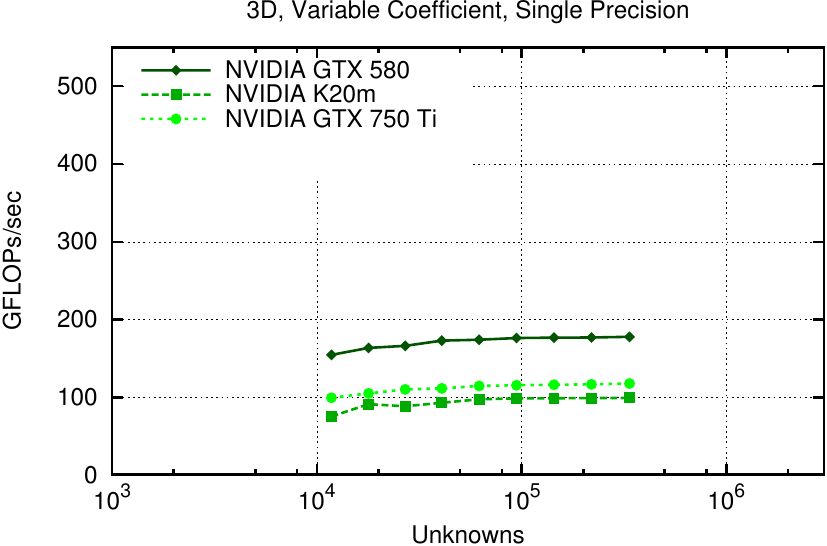}} \hfill
  \subfigure[3D, double precision]{\includegraphics[width=0.48\textwidth]{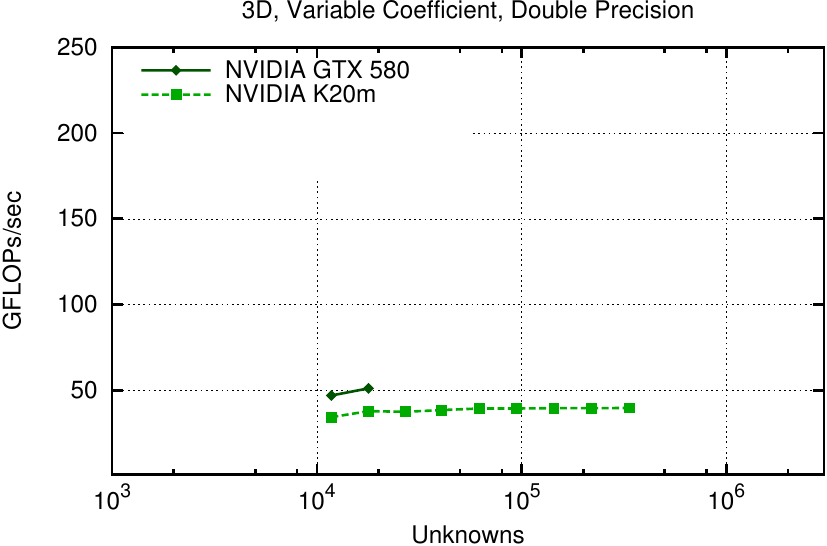}}
  \caption{FLOP rates for the different GPUs considered in this benchmark for evaluating the residual for the linear elasticity equations on unstructured meshes of the the unit square (2d) and the unit cube (3d).}
  \label{fig:flopElasticity}
\end{figure}

Fig.~\ref{fig:flopElasticity} depicts the benchmark results obtained for solving the linear elasticity equations.
The vector-valued basis reduces the algorithmic balance $\beta$ by roughly two-fold compared to the Poisson equation.
This is also reflected in the observed performance on a GTX 580: 
In two dimensions, $220$ GFLOP/sec in single precision ($300$ GFLOP/sec for Poisson) and $80$ GFLOP/sec in double precision ($150$ GFLOP/sec for Poisson).
In three dimensions, we obtained $180$ GFLOP/sec in single precision ($420$ GFLOP/sec for Poisson) and $70$ GFLOP/sec in double precision ($150$ GFLOP/sec for Poisson).

\subsection{Reproducing the Results}\label{sec:reproduce}

All the foregoing examples can be run using the PETSc libraries.
You will need to configure with OpenCL support, for example
\begin{lstlisting}[language=sh,basicstyle=\footnotesize\tt]
./configure --with-opencl-include=/path/to/CL
  --with-opencl-lib=/path/to/libOpenCL.so --download-mpich
  --download-triangle --download-ctetgen
\end{lstlisting}

The data for the Poisson problem was collected using multiple runs of the form
\begin{lstlisting}[language=sh,basicstyle=\footnotesize\tt]
 ./ex12 -petscspace_order 1 -run_type perf -variable_coefficient field
  -refinement_limit 0.00001 -show_solution false -petscfe_type opencl
  -petscfe_num_blocks 16 -petscfe_num_batches 8
  -petscfe_opencl_real_type double
\end{lstlisting}
where \lstinline|-petscfe_num_blocks| specifies the number of blocks per patch and \lstinline|petscfe_num_batches| selects the number of blocks per batch.
The option \lstinline|-refinement_limit| is passed to the mesh generator (triangle for triangular meshes, tetgen for tetrahedral meshes) and provides control over the problem size.

\section{Conclusions}

We have produced a high performance implementation of FEM quadrature for low order elements, amenable to high throughput
architectures such as GPUs. The performance gain is due to flexible vectorization without a reduction over threads, made
possible by the thread transposition construct we introduced. The OpenCL implementation has been integrated into
PETSc, showing the ease of implementation and providing an open testbed for further work. In future work, we will
integrate this work into existing application, such as the PyLith code for crustal deformation~\cite{AagaardKnepleyWilliams13}.

\appendix

Below we give the overall structure of the integration code, eliminating extraneous detail such as variable declaration
and precise indices. The full code can be found in the PETSc distribution \href{https://bitbucket.org/petsc/petsc/src/2d8c9f8046ca79ac5ca658baa3f0377fa9aa8472/src/dm/dt/interface/dtfe.c?at=master#cl-2879}{OpenCL implementation}.
\begin{alltt}\scriptsize
// N_cb  Number of serial cell batches
// N_bl  Number of concurrent blocks
void integrateElementQuadrature(int N_cb, realType *coefficients,
    realType *jacobianInverses, realType *jacobianDeterminants,
    realType *elemVec)
\{
  int dim    = spatialDim;         // Spatial dimensions
  int N_b    = numBasisFunctions;  // Basis functions
  int N_comp = numBasisComponents; // Basis function components
  int N_q    = numQuadPoints;      // Quadrature points
  int N_bt   = N_b*N_comp;         // Total scalar basis funcs
  int N_bst  = N_bt*N_q;           // Block size
  int N_t    = N_bst*N_bl;         // Threads
  int N_bc   = N_t/N_comp;         // Cells/batch
  int N_c    = N_cb * N_bc;        // Total cells
  int N_sbc  = N_bst/(N_q*N_comp); // Serial basis cells
  int N_sqc  = N_bst/N_bt;         // Serial quad cells

  /* Load quadrature weights */
  w = weights_0[q];
  /* Load basis tabulation phi_i for this cell */
  phi_i[q,b]    = Basis_0[q,b];
  phiDer_i[q,b] = BasisDerivatives_0[q,b];

  for (int batch = 0; batch < N_cb; ++batch) \{
    /* Load geometry */
    detJ[c] = jacobianDeterminants[c];
    invJ[c] = jacobianInverses[c];
    /* Load coefficients u_i for this cell */
    u_i[c] = coefficients[c];

    /* Map coefficients to values at quadrature points */
    for (int c = 0; c < N_sqc; ++c) \{
      u     = 0.0; // u(x_q), Value of the field at x_q
      gradU = 0.0; // du/dx(x_q), Value of the gradient at x_q
      /* Get field and derivatives at this quadrature point */
      for (int i = 0; i < N_b; ++i) \{
        for (int comp = 0; comp < N_comp; ++comp) \{
          u[comp]     += u_i[c,b]*phi_i[q,b];
          gradU[comp] += u_i[c,b]*(invJ[c] * phiDer_i[q,b]);
        \}
      \}
      /* Process values at quadrature points */
      f_0[c,q,comp] = f0_func(u, gradU, c)*detJ[c]*w;
      f_1[c,q,comp] = f1_func(u, gradU, c)*detJ[c]*w;
    \}

    /* ==== TRANSPOSE THREADS ==== */
    syncronize_threads();

    /* Map values at quadrature points to coefficients */
    for (int c = 0; c < N_sbc; ++c) \{
      e_i = 0.0;
      for (int q = 0; q < N_q; ++q) \{
        e_i += phi_i[q,b]*f_0[c,q,comp];
        e_i += invJ * phiDer_i[q,b] * f_1[c,q,comp];
      \}
      /* Write element vector for N_cbc cells at a time */
      elemVec[c,b] = e_i;
    \}
  \}
\}
\end{alltt}

\section*{Thanks} We would like to thank Hans Petter Langtangen for organizing a visit to Simula Research which was the
genesis of this paper.
%MGK acknowledges partial support from DOE Contract DE-AC02-06CH11357 and NSF Grant OCI-1147680.